\documentclass[twocolumn,aps,prl,showpacs]{revtex4}

\usepackage{epsfig}

\begin{document}

\title{Molecular Nanostructures on the Surface of a $d_{x^2-y^2}$-Superconductor}
\author{Roy H. Nyberg and Dirk K.~Morr}
\affiliation{Department of Physics, University of Illinois at Chicago, Chicago, IL 60607}
\date{\today}

\begin{abstract}

We study molecular nanostructures on the surface of a
$d_{x^2-y^2}$-wave superconductor. We show that the interplay
between the molecular nanostructure's internal excitation spectrum
and quantum interference of the scattered host electrons leads to a
series of novel effects in the local density of states. We
demonstrate that these effects give insight into both the nature of
the superconducting pairing correlations and of the intermolecular
interactions.

\end{abstract}

 \maketitle

Nanostructures provide the intriguing possibility of manipulating in
a controlled way the electronic structure of the host system they
reside on \cite{nanoexp}. This level of control opens new venues for
studying the complex electronic structure of many strongly
correlated electron systems. For example, it was suggested that
nanostructures can elucidate the nature of superconducting
correlations in conventional \cite{Morr04} and unconventional
superconductors \cite{2imp}. Nanostructures formed from more complex
building blocks, such as molecules with internal degrees of freedom,
provide a novel way of manipulating and simultaneously gaining
insight into the electronic structure of complex systems. The
internal vibrational and rotational excitations of single molecules
have been intensively studied by inelastic tunneling spectroscopy
\cite{singlemol,review}. Their interaction with and effect on the
electronic structure of a metallic surface was recently investigated
by Gross {\it et al.}~\cite{Gro04}.

In this Letter we study nanostructures composed of two molecules
that reside on the surface of an unconventional
$d_{x^2-y^2}$-superconductor that represents the family of
high-temperature superconductors (HTSC). We show that the interplay
between the molecular nanostructure's internal excitation spectrum
and quantum interference of the scattered host electrons leads to a
series of novel effects in the local density of states (LDOS) of the
superconductor. Our main results are threefold. First, we identify
three types of intermolecular interactions and show that each of
them leads to {\it qualitatively} distinguishable features in the
LDOS. This, in turn, allows one to identify the nature of
intermolecular interaction from experimental measurements of the
LDOS. Second, we demonstrate that the LDOS changes with the
molecules' distance and orientation relative to the underlying host
lattice. This effect permits us to probe the spatial dependence of
superconducting correlations. Third, we show that by exciting
specific energy levels of the molecular structure, the LDOS can be
manipulated in a controlled manner. These results intricately relate
the study of strongly correlated electron systems with the further
development of molecular electronics.

A nanostructure consisting of $N$ molecules, represented by local
bosonic modes \cite{Hew79}, on the surface of a two-dimensional
$d_{x^2-y^2}$ superconductor possesses the Hamiltonian
$H=H_e+H_b+H_{int}$ where
\begin{eqnarray}
&H_e&=\sum_{\bf{k}, \sigma=\uparrow,\downarrow} \epsilon_{\bf{k}}
c^{\dagger}_{\mathbf{k,\sigma}}
c_{\mathbf{k,\sigma}} + \sum_{\mathbf{k}} \Delta _{\mathbf{k}}c_{\mathbf{k}%
,\uparrow}^{\dagger }c_{\mathbf{-k},\downarrow}^{\dagger} + h.c. \ , \nonumber \\
&H_b&= \omega_{\mathbf{0}} \sum_{i} a^{\dagger}_{\mathbf{r}_i} a_{%
\mathbf{r}_i} + J \sum_{i,j} \Psi \big[ a^{\dagger}_{{\bf r}_i},
a_{{\bf r}_i}, a^{\dagger}_{{\bf r}_j},
a_{{\bf r}_j} \big] \ , \nonumber \\
&H_{int}&= g \sum_{i, \sigma} \left(a^{\dagger}_{\mathbf{r}_i} +
a_{\mathbf{r}_i} \right) c^{\dagger}_{\mathbf{r}_i,\sigma}
c_{\mathbf{r}_i,\sigma} \ . \label{HG}
\end{eqnarray}
$H_e$ and $H_b$ describe the unperturbed superconductor and
nanostructure, respectively, and $H_{int}$ represents the
interaction between them. $c^\dagger, a^\dagger$ are the fermionic
and bosonic creation operators, respectively. $\Delta
_{\mathbf{k}}=\Delta_0 (\cos k_x - \cos k_y)/2$ is the
$d_{x^2-y^2}$-wave superconducting gap and $\epsilon_{\bf k}=-2t
\left( \cos k_x + \cos k_y \right) - 4 t' \cos k_x \cos k_y - \mu$
is the normal state tight-binding dispersion. $\omega_0$ is the
characteristic frequency of the $N$ modes located at ${\bf r}_i$
($i=1,..,N$). $\Psi$ is a quadratic functional in the bosonic
operators and describes the interaction between modes at sites ${\bf
r}_i$ and ${\bf r}_j$ with strength $J$ (three specific forms of
$\Psi$ are discussed below). $g$ is the boson-fermion scattering
vertex. The fermionic Greens function of the unperturbed (clean)
system in Nambu notation is
\begin{equation}
\hat{G}^{-1}_0(\mathbf{k},i\omega_n)=\left[ i\omega_n \tau_0 - \epsilon_{%
\mathbf{k}} \tau_3 \right] \sigma_0 + \Delta_{\mathbf{k}} \tau_2
\sigma_2 \ , \label{G0}
\end{equation}
where $\sigma_i$ and $\tau_i$ are the Pauli matrices in spin and
Nambu space, respectively. Diagonalizing $H_b$ via a Bogoliubov
transformation to new operators $b^\dagger_l, b_l$ ($l=1,...,N$),
one obtains $H_{b}=\sum_{l} \Omega_l b^\dagger_l b_l$ and
\begin{equation}
H_{int}=\sum_{i,l,\sigma} g_{{\bf r}_i}^{(l)} \left(
b^\dagger_l + b_l \right) c^{\dagger}_{\mathbf{r}_i,\sigma} c_{%
\mathbf{r}_i,\sigma} \ ,
\end{equation}
where $g_{{\bf r}_i}^{(l)}$ is a site and mode dependent interaction
vertex, and $\Omega_l$ is the bosonic energy spectrum. The full
fermionic Green's function is now given by
\begin{eqnarray}
\hat{G}({\bf r},{\bf r}^{\prime}, \omega_n)&=&\hat{G}_0({\bf
r},{\bf r}^{\prime},\omega_n)
+ \sum_{i,j} \hat{G}_0({\bf r},{\bf r}_i,\omega_n)  \nonumber \\
& & \times \hat{\Sigma}({\bf r}_i,{\bf r}_j,\omega_n) \hat{G}_0({\bf
r}_j,{\bf r}^{\prime},\omega_n) \label{fullG} \ ,
\end{eqnarray}
where $\hat{\Sigma}$ is the full fermionic self-energy obtained from
the Bethe-Salpeter equation
\begin{eqnarray}
\hat{\Sigma}({\bf r}_i, {\bf r}_j, \omega_n)&=&\hat{\Sigma}_0({\bf
r}_i, {\bf r}_j, \omega_n) +\sum_{p,q} \hat{\Sigma}_0({\bf r}_i,{\bf r}_p,\omega_n)  \nonumber \\
& & \times \hat{G}_0({\bf r}_p,{\bf r}_q, \omega_n)
\hat{\Sigma}({\bf r}_q, {\bf r}_j,\omega_n) \label{BeSa}
\end{eqnarray}
with $p,q=1,...,N$ and
\begin{eqnarray}
\hat{\Sigma}_0({\bf r}_i,{\bf r}_j,\omega_n)&=&T \sum_{m,l} \tau_3 g_{{\bf r}_i}^{(l)} {%
\hat G}_0({\bf r}_i,{\bf r}_j,\omega_{n}+\nu_{m}) \nonumber \\
& & \times D_{l}(\nu_{m}) \tau_3 g_{{\bf r}_j}^{(l)} \ .
\label{Sigma0}
\end{eqnarray}
The retarded form of the bosonic propagator is given by
\begin{equation}
D^R_{l}(\omega)= \left[\omega+i\Gamma(\omega)-\Omega_l
\right]^{-1} - \left[\omega+i\Gamma(\omega)+\Omega_l\right]^{-1} \
. \label{prop}
\end{equation}
The lowest order vertex corrections scale as $\delta g/g \sim
(gk_F/4 v_F)^2 F$ where $F$ is a function of $O(1)$ for bosonic and
fermionic frequencies smaller than $\Delta_0$ \cite{Morr03}. For the
parameter range considered below, $gk_F/4v_F \ll 1$ and vertex
corrections can thus be neglected. The interaction, $H_{int}$, not
only leads to changes in the fermionic LDOS, but also affects the
bosonic excitation spectrum twofold. First, $H_{int}$ shifts the
unperturbed bosonic frequencies, a shift that we take to be included
in the effective values for $\Omega_l$ considered below. Second, the
bosonic modes acquire a finite lifetime, $\Gamma(\omega) \not = 0$.
The lowest order bosonic self-energy correction yields
$\Gamma(\omega)=g^2 \omega^3 /(6 \pi v_F^2 v_\Delta^2)$, where $v_F$
($v_\Delta$) is the Fermi (superconducting) velocity at the nodal
points perpendicular (parallel) to the Fermi surface. Since for all
cases considered below $\Gamma(\omega_0) \ll \omega_0$, which
requires $g \ll 50 \Delta_0$, $\Gamma(\omega)$ can be neglected and
we take for simplicity $\Gamma(\omega)=0^+$. Finally, the LDOS,
$N=A_{11}+A_{22}$ with $A_{ii}({\bf r},\omega)=-2{\rm Im}\,
\hat{G}_{ii}({\bf r},\omega+i\delta)$ is obtained numerically from
Eqs.(\ref{fullG}) with $\delta=0.2$ meV.

We showed in Ref.~\cite{Morr03} that a single local mode on the
surface of a $d_{x^2-y^2}$-superconductor is pair-breaking and
induces a fermionic resonance state, whose spectroscopic signature
are two peaks in the LDOS. Introducing a second mode leads to two
new physical effects that introduce {\it qualitative} changes in the
LDOS. First, due to quantum interference of the scattered electrons,
the calculation of the LDOS involves not only the local Greens
function, $\hat{G}_0({\bf r}_i, {\bf r}_j)$, and self-energy,
$\hat{\Sigma}({\bf r}_i, {\bf r}_j)$, with ${\bf r}_i = {\bf r}_j$,
but also the non-local ones with ${\bf r}_i \not = {\bf r}_j$. As a
result, the LDOS is determined by the distance and orientation of
the modes relative to the lattice, which is described by $\Delta
{\bf r}={\bf r}_i - {\bf r}_j=(n_x, n_y )$ (we set the lattice
constant $a_0=1$). Second, the intermolecular interaction $\Psi$
leads to characteristic features in the LDOS that, as we show below,
depend on the specific form of $\Psi$.

\begin{figure}[h]
\epsfig{file=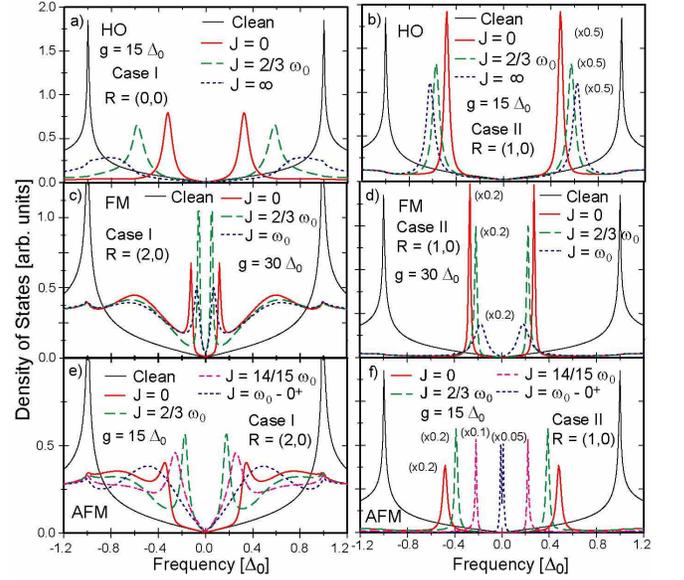,width=8.5cm} \caption{LDOS at ${\bf R}$ as a
function of frequency for $\omega_0=0.6 \Delta_0$ and three
intermolecular interactions: (a),(b) HO coupling, (c),(d) FM
coupling, and (e),(f) AFM coupling. Scaling of the curves is
indicated.} 
\label{Tzero}
\end{figure}

In order to distinguish between the effects of quantum interference,
superconducting correlations and inter-molecular interaction on the
LDOS, we consider first a simplified electronic bandstructure with
$t=300$ meV, $\mu=t^\prime=0$ and set $J=0$. In this case,
superconducting correlations, as represented by the non-vanishing of
the non-local anomalous Greens function, $F_{\Delta r}$, and
self-energy, $\Phi_{\Delta r}$ (the off-diagonal elements of
$\hat{G}_0$ and $\hat{\Sigma}_0$, respectively), are present for
$(n_x+n_y) {\rm mod } 2=1$ ({\it case I}) and vanish otherwise ({\it
case II}). We find that the presence of superconducting correlations
leads to qualitative differences in the LDOS between case I and II
that are {\it universal}, i.e., independent of the specific
realization of case I or II considered. For definiteness, we
consider below two modes located at ${\bf r}_1=(0,0)$ and ${\bf
r}_2=(1,0)$  (case I) and ${\bf r}_1=(0,0)$ and ${\bf r}_2=(2,0)$
(case II) and present the resulting LDOS at $T=0$ in
Fig.~\ref{Tzero}(a) and (b).
%
%
The coupling of the resonance states due to quantum interference
only (see curves for $J=0$) leads to the formation of bonding and
antibonding states and an energy splitting between them. Only one of
these states is located at energies smaller than $\Delta_0$,
resulting in a particle-like and hole-like peak in the LDOS. Quantum
interference leads to oscillations in the frequency of the resonance
peaks when $\Delta {\bf r}$ is changed (cf.~Figs.~\ref{Tzero}(a) and
(b)), similar to the case of non-magnetic impurities \cite{2imp}.
Next, we discuss the effects of three different intermolecular
interactions on the frequency and shape of the resonance peaks.

\indent \textit{Harmonic Oscillator (HO) Coupling}: A coupling of
the bosonic modes via a quadratic potential is the quantum
mechanical analog of two harmonic oscillators coupled by a spring.
In this case \cite{Cohen} $\Psi= \left( b^{\dagger}_{\bf r_1} +
b_{\bf r_1} - b^{\dagger}_{\bf r_2} - b_{\bf r_2} \right)^2$ with $J
\geq 0$. After diagonalization, one has $\Omega_1=\omega_0$,
$\Omega_2=\omega_0 \sqrt{1 + 4 \lambda}$ with $\lambda \equiv
J/\omega_0$, $g_{\mathbf{r}_{1,2}}^{(1)}=g/\sqrt{2}$ and
$g_{\mathbf{ r}_{1,2}}^{(2)}=\pm g/[ \sqrt{2} (1+4 \lambda)^{1/4}
]$.  The resulting LDOS for several values of $J$ is shown in
Fig.~\ref{Tzero}(a) and (b) for cases I and II, respectively. With
increasing $J$, $\Omega_2$ increases while $|g_{{\bf
r}_{1,2}}^{(2)}|$ decreases, the electronic scattering hence becomes
weaker and the resonance peaks are shifted to higher energies. In
case I, the peaks move close to $\pm \Delta_0$ as $J$ increases from
$J=0$ to $J=\infty$, an effect that is independent of $g$ and
directly related to a non-zero $F_{\Delta r}$ and $\Phi_{\Delta r}$.
In contrast, in case II, the frequency shift is much smaller. For $J
= \infty$, only the bosonic {\it in-phase} mode couples to the
fermionic system. This coupling is much weaker in case I than in
case II due to the superconducting correlations.

\textit{Ferromagnetic (FM) Coupling}: A second type of
intermolecular interaction arises if each molecule possesses
$(2L+1)$ states that are represented by pseudo-spins ${\bf
L}_{1,2}$, whose interaction is given by the anisotropic Heisenberg
Hamiltonian
\begin{equation}
H_{spin}=J_z\, L_1^z L_2^z + J_\pm \left( L_1^x L_2^x + L_1^y L_2^y
\right)  \ .  \label{Hspin1}
\end{equation}
For $J_z<0$ and $|J_z|>|J_\pm|$, the pseudo-spins are
ferromagnetically aligned (i.e., the molecules prefer to be in the
same state). Performing a Holstein-Primakoff transformation followed
by a large-$L$ expansion up to order $O(L)$, we obtain $H_b$ in
Eq.(\ref{HG}) with $\omega_0 = - J_zS>0$, $J=J_\pm S \geq 0$ and
$\Psi=b^\dagger_{\bf r_1} b_{\bf r_2}+b^\dagger_{\bf r_2} b_{\bf
r_1}$. After diagonalization, one has $\Omega_{1,2}=\omega_0 \mp J$,
$g_{\bf r_{1,2}}^{(1)}=\mp g/\sqrt{2}$ and $g_{\bf
r_{1,2}}^{(2)}=g/\sqrt{2}$. The operators $b^\dagger_{\bf r_i},
b_{\bf r_i}$ represent the excitations of the pseudo-spin dimer
which scatter fermions via $H_{int}$. For $J_z<J_\pm$, i.e.,
$J>\omega_0$, the system becomes unstable since the ferromagnetic
alignment of the pseudo-spins is destroyed.

For small values of $g$, the resonance peaks move towards lower
energies in cases I and II with increasing $J$ (not shown). However,
for intermediate $g \approx 30 \Delta_0$, this behavior changes
qualitatively for case I; the resonance frequencies exhibit a
non-monotonic dependence on $J$ [Fig.~\ref{Tzero}(c)], in contrast
to case II [Fig.~\ref{Tzero}(d)]. For even larger $g$, the resonance
peaks shift to higher energies in case I, and to lower energies in
case II with increasing $J$. A detailed analysis shows that this
qualitative difference arises from a non-zero $F_{\Delta r}$ and
$\Phi_{\Delta r}$ in case I. Note, in case II (Fig.~\ref{Tzero}(d)),
the width of the resonance peaks increases when it shifts to lower
energies, in contrast to the effect expected from a decreasing
residual DOS \cite{2imp}. Here, however, the width of the resonance
peaks is also determined by the imaginary part of the self-energy
which vanishes for $|\Omega| < \Omega_{1}$. Since $\Omega_{1}$
decreases with increasing $J$, the resonance state becomes more
strongly damped once $\Omega_{1}<\omega_{res}$. In
Fig.~\ref{Tzero}(d), we have $\Omega_{1}=\omega_0>\omega_{res}$ for
$J=0$, but $\Omega_{1}<\omega_{res}$ for $J=2\omega_0/3$
($\Omega_{1}=0.2 \Delta_0$) and $J=\omega_0$ ($\Omega_{1}=0$),
resulting in an increased peaks' width for the latter two values of
$J$.

\textit{Antiferromagnetic (AFM) Coupling}: For antiferromagnetic
alignment of the pseudo-spins ($J_z>0$, $|J_z|>J_\pm>0$ in
Eq.(\ref{Hspin1})), the molecules prefer to be in ``opposite"
states. Performing a Holstein-Primakoff transformation followed by a
large-$L$ expansion, one obtains $\Psi=b^{\dagger}_{{\bf r}_1}
b^{\dagger}_{{\bf r}_2} + b_{{\bf r}_1} b_{{\bf r}_2}$ resulting in
two degenerate energy dispersions $\Omega_{1,2}=\omega_0
\sqrt{1-\lambda^2}$ with $\lambda \equiv J/\omega_0$,
$g_{\mathbf{r}_1}^{(1,2)}= \pm g u_{1,2}$,
$g_{\mathbf{r}_2}^{(1,2)}= \mp g u_{2,1}$ and
$u^2_{1,2}=[1/\sqrt{1-\lambda^2} \pm 1]/2$. For $J>\omega_0$, i.e.,
$|J_z|<J_\pm$, the system becomes unstable since the
antiferromagnetic alignment of the molecular pseudo-spins is
destroyed.

With increasing $J$, $\Omega_{1,2}$ decreases while
$g_{\mathbf{r}_2}^{(1,2)}$ increases, thus leading to an increase in
the scattering strength. As a result, the resonance peaks shift
monotonically to lower energies in case II [see
Fig.~\ref{Tzero}(f)], while their width decreases since
$\omega_{res}<\Omega_{1,2}$ for all $J$. In the limit $J \rightarrow
\omega_0$, $g_{\mathbf{r}_2}^{(1,2)}$ diverge, and the scattering
becomes unitary. Since the local and non-local ${\hat G}_0$ and
${\hat \Sigma}$ vanish for $\omega = 0$, the resulting resonance
peaks are located at zero energy and the LDOS vanishes at the
molecules' sites. In contrast, in case I [Fig.~\ref{Tzero}(e)], the
resonance peaks first shift to lower energies with increasing $J$,
but eventually move back to higher energies and broaden. This
behavior arises from the nonvanishing of the non-local Greens
function and self-energy for $\omega = 0$.

\begin{figure}[h]
\epsfig{file=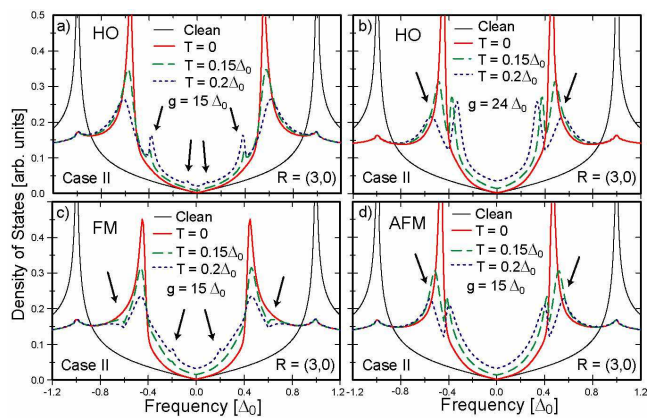,width=8.5cm} \caption{ Temperature dependence
of the LDOS for $\omega_0=0.6\Delta_0$ and $J=2 \omega_0/3$: (a),
(b) HO-coupling, (c) FM-coupling, and (d) AFM-coupling.}
\label{DOStemp}
\end{figure}

At $T=0$, the imaginary part of the normal self-energy (the diagonal
element of $\hat{\Sigma}_0$) possesses logarithmic divergences that
lead to dips in the LDOS at $\pm \omega_{1,2}^+$
($\omega_{1,2}^{\pm}= \Delta_0 \pm \Omega_{1,2}$)
\cite{Morr03,Bal02}. At $T \not = 0$, the positive energy levels of
the bosonic modes, represented by the first term in Eq.(\ref{prop}),
become populated, opening a new channel for fermionic scattering. As
a result, $\hat{\Sigma}_0$ acquires logarithmic divergences at $\pm
\omega_{1,2}^{-}$. This leads to either peaks or dips in the LDOS,
depending on whether $\omega_{i}^{-}$ is smaller or larger than the
frequency of the resonance peaks, $\pm \omega_{res}$. The LDOS for
$T \not = 0$ is shown in Fig.~\ref{DOStemp}.
In Figs.~\ref{DOStemp}(a) and (b), we plot the LDOS for HO coupling
in case II for $g=15 \Delta_0$ and $g=24 \Delta_0$, respectively.
For $g=15 \Delta_0$, $\omega_{1,2}^{-}<\omega_{res}$ and the
logarithmic divergences lead to four peaks (indicated by arrows) in
the LDOS. Their intensity is governed by $n_B( \Omega_{i})$ and thus
larger at $\pm \omega_{1}^{-}$ than at $\pm \omega_{2}^{-}$.
Simultaneously, opening of the new scattering channels leads to an
increase in the imaginary part of the self-energy, resulting in a
shift of the resonance peaks to higher energies and an increase in
their width. In contrast, for $g=24 \Delta_0$, $\omega_{1}^{-}
\approx \omega_{res}$, leading to a splitting of the resonance peaks
and to a dip in the LDOS. The second set of divergences at
$\omega_{2}^{-}$ induces two peaks in the LDOS, which due to the
overall increase in the LDOS at lower energies are barely
perceptible. The LDOS for FM coupling and AFM coupling in case II is
shown in Fig.~\ref{DOStemp}(c) and (d), respectively. For FM
coupling, the LDOS exhibits two peaks and two dips since
$\omega_{1}^{-}<\omega_{res}<\omega_{2}^{-}$. In contrast, for AFM
coupling, the two modes are degenerate and since
$\omega_{1,2}^{-}>\omega_{res}$, the logarithmic divergences result
in a single set of dips. The LDOS in case I (not shown) exhibits
qualitatively similar behavior to the ones discussed above for all
three couplings. Note that the frequencies of the features arising
from the logarithmic divergences are {\it independent} of $g$, which
thus permits a direct measurement of $\pm \omega_{1,2}^{\pm}$.

The degeneracy of the modes for AFM coupling results in a LDOS with
only two features at $\pm \omega_{1}^{-}$ and $\pm \omega_{1}^{+}$,
respectively, in {\it qualitative} contrast to the LDOS for FM and
HO coupling. The latter two couplings, however, can be distinguished
by comparing $\omega_0$ obtained from the LDOS near a single
molecule \cite{Morr03} with $\omega_{1,2}^{\pm}$. Thus the nature of
the intermolecular interaction, similar to a molecule's internal
structure \cite{Gro04}, can be identified experimentally from the
specific features it induces in the LDOS.

%
%
\begin{figure}[t]
\epsfig{file=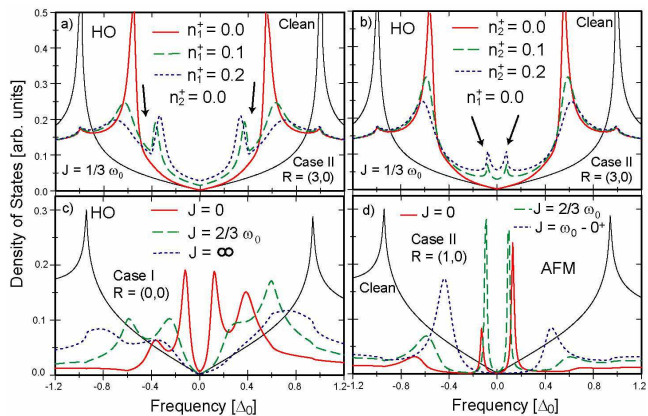,width=8.5cm} \caption{LDOS at $T=0$ for
$\omega_0=0.6\Delta_0$ and $g=15 \Delta_0$. (a),(b) The positive
energy branches of the bosonic modes are selectively populated for
HO-coupling with $n_{1,2}^{-}=1$ and $J=\omega_0/3$. (c), (d) LDOS
for a HTSC band structure and (c) HO-coupling, and (d)
AFM-coupling.} \label{Fpump}
\end{figure}
Changing the population, $n_{1,2}^{\pm}$, of the energy levels $\pm
\Omega_{1,2}$, for example by optical means \cite{optical}, opens
intriguing venues to manipulate the local electronic structure of
the host material. In Fig.~\ref{Fpump}(a),(b), we present the LDOS
for HO coupling with $n_{1,2}^{-}=1$ and $n_{1,2}^{+} \not = 0$. A
non-zero $n_{1}^{+}$ [Fig.~\ref{Fpump}(a)] induces dips in the LDOS
at $\pm \omega_{1}^{-}$, while an increase of $n_{2}^+$
[Fig.~\ref{Fpump}(b)] leads to peaks at $\pm \omega_{2}^{-}$,
similar to the temperature induced changes discussed above.

Finally, the use of a bandstructure representative of the HTSC with
$t^\prime/t=-0.4$ and $\mu/t=-1.18$ \cite{Dam03} mixes the effects
of intermolecular interaction, superconducting correlations and
quantum interference on the LDOS. For example, the $J$-dependence of
the LDOS for HO coupling [Fig.~\ref{Fpump}(c)] is similar to the one
shown in Fig.~\ref{Tzero}(a), while the LDOS for AFM coupling
[Fig.~\ref{Fpump}(d)] shows a dependence on $J$ that is
qualitatively different from that in Fig.~\ref{Tzero}(f).

In conclusion, we study the effects of molecular nanostructures on
the electronic structure of a $d_{x^2-y^2}$-wave superconductor. We
show that different intermolecular interactions lead to {\it
qualitatively} distinguishable signatures in the LDOS. By changing
the population of the bosonic excitations, one can manipulate the
host's electronic structure and gain further insight into the nature
of unconventional superconducting correlations.

We would like to thank J.C. Davis, T. Imbo, and K.-H. Rieder for
stimulating discussions. D.K.M. acknowledges support from the
Alexander von Humboldt foundation.

\end{document}